\documentstyle[aps,epsf]{revtex}
\newcommand{\be}{\begin{equation}}
\newcommand{\ee}{\end{equation}}
\newcommand{\bea}{\begin{eqnarray}}
\newcommand{\eea}{\end{eqnarray}}
\newcommand{\nn}{\nonumber \\}
\newcommand{\eqn}[1]{(\ref{#1})}
\font\mybb=msbm10 at 10pt
\def\bb#1{\hbox{\mybb#1}}

\def\bR {\bb{R}}
\def\bE {\bb{E}}

\def\pa {\partial}
\def\sac{\, , \qquad}
\newcommand{\bfna}{\mbox{\boldmath $\nabla$}}
\newcommand{\bfa}{\mbox{\boldmath $A$}}
\newcommand{\un}{\underline}

\begin{document}

\twocolumn[\hsize\textwidth\columnwidth\hsize\csname
@twocolumnfalse\endcsname

\rightline{DAMTP-2001-21}
\rightline{hep-th/0103030}

\title{Supertubes} 
\author{David Mateos and Paul K. Townsend\\}
\address{
DAMTP, Centre for Mathematical Sciences, \\
University of Cambridge, Wilberforce Road, \\
Cambridge CB3 0WA,  UK \\
{\small D.Mateos@damtp.cam.ac.uk, P.K.Townsend@damtp.cam.ac.uk} \\
}
\maketitle
\begin{abstract}
It is shown that a IIA superstring carrying D0-brane charge can be
`blown-up', in a {\it Minkowski vacuum} background, to a 
(1/4)-supersymmetric tubular D2-brane, supported against collapse 
by the angular momentum generated by crossed electric and magnetic 
Born-Infeld fields. This `supertube' can be viewed as a worldvolume 
realization of the sigma-model Q-lump.
\end{abstract}
\vskip2pc]

\section{Introduction}

M-theory can be viewed as an extension of string theory to a
`democratic' theory of branes, with dualities that allow the whole
theory to be constructed from any brane. One aspect of this is that
a given brane may often be considered as a partially collapsed 
version of another, higher-dimensional, brane. Conversely, there may be
circumstances in which the lower-dimensional brane is `blown-up' into the  
higher-dimensional one. One example of this is the observation 
that a IIA superstring can be blown-up to a tubular D2-brane
by placing it in an appropriate (non-supersymmetric) background \cite{E};
the background fields impose an external force that prevents the collapse
of the D2-brane. Another example is the observation that a similar
non-trivial background can blow-up a collection of D0-branes into a
fuzzy 2-sphere \cite{M}; this can again be considered as a D2-brane 
prevented from collapse by an external force \cite{SJR}. Another way
to support a brane against collapse to a lower-dimensional one 
is to allow it to carry angular momentum; the case of branes on 
spheres, first analyzed in \cite{HN}, has recently found an
application to `giant gravitons' \cite{MST}; these partially preserve 
the supersymmetry of an $AdS_n\times S^m$ supergravity 
background \cite{goliath}, a fact that is presumably related to the
appearance of angular momentum in the anticommutator of
$AdS$-supersymmetry charges.  

Here we shall show that a cylindrical, or tubular, D2-brane in a 
{\it Minkowski vacuum} spacetime can be supported against collapse by 
the angular momentum generated by crossed electric and magnetic 
Born-Infeld (BI) fields. The construction involves a uniform electric
field along the tube, and a constant magnetic flux. The electric field
can be interpreted as a `dissolved' IIA superstring, so the
tube is a IIA superstring that has been `blown-up' to a tubular D2-brane.
The magnetic flux can be interpreted as a `dissolved' D0-brane charge (per
unit length). There are thus similarities to some of the previously
proposed stabilization mechanisms but also some differences. Firstly,
the background in our case is trivial; there is {\it no external
force}. Secondly, the tubular D2-brane configuration presented here 
is {\it supersymmetric}; it preserves 1/4 of the supersymmetry of 
the IIA Minkowski vacuum, hence the term `supertube'.

At first sight it would appear unlikely that a brane configuration
stabilized by angular momentum could be supersymmetric. For a start,
supersymmetry requires a time-independent energy density, which is
certainly non-generic for configurations with non-zero angular
momentum. Even if this can be arranged, there is still the fact that
a supersymmetric configuration in Minkowski space minimizes the energy
subject to fixed values of the central charges appearing in the 
supersymmetry algebra, but {\it angular momentum is not one of these
charges}. However, although these considerations may make supersymmetric 
stabilization by angular momentum unlikely, they do not make it 
impossible. Indeed, in the somewhat different
context of field theory solitons there are {\it known} examples of
supersymmetric solitons that are stabilized by angular momentum.
One such example is the $Q$-lump of certain massive $D=3$ supersymmetric
sigma-models \cite{L}, and the D2-brane supertube is essentially its
{\it worldvolume} realization.

The $Q$-lump saturates an energy bound of the form $M= |L| + |Q|$, where $L$
is a topological (lump) charge and $Q$ is a Noether charge \cite{A}. 
When viewed as a string-like solution of the maximally supersymmetric 
$D=4$ sigma-model, it preserves 1/4 of the supersymmetry of the 
sigma-model vacuum \cite{GPTT}.
The simplest $Q$-lump-string has cylindrical symmetry \cite{footnote},
but for large $Q$ its energy density is {\it not}
concentrated on the axis of symmetry but rather in a hollow tube of radius
$Q$. Therefore the $Q$-lump-string tube can be interpreted, for large $Q$,
as a lump-string that has been `blown-up' into a cylindrical tube of
kink-2-brane
\cite{AT}. This tube is prevented from collapse by a centrifugal force,
generated by an angular momentum proportional to the charge $Q$. Because
the effective action for the kink 2-brane is a Dirac-Born-Infeld (DBI)
action
\cite{GPTT}, there should be a {\it worldvolume} description of the
`blown-up' sigma-model lump-string as a 1/4-supersymmetric solution 
of the $D=4$ DBI equations. 
Any such solution will also solve the DBI equations of the $D=10$
IIA D2-brane (in a IIA Minkowski vacuum) and must preserve at least two
supersymmetries (this being 1/4 of the eight supersymmetries of the 
sigma-model vacuum). This solution is precisely the D2-brane 
`supertube' to be discussed below;
it actually preserves eight of the thirty-two supersymmetries of the IIA
vacuum and is therefore still 1/4-supersymmetric.

\section{Energetics of the Supertube}
The D2-brane Lagrangian, for unit surface tension, is
\be
{\cal L} \equiv - \Delta = -\sqrt{\det (g +F)} \,,
\ee
where $g$ is the induced worldvolume 3-metric and $F$ is the BI 2-form
field strength. We shall choose spacetime coordinates such that the $D=10$
Minkowski metric is
\be
ds^2 = -dT^2 + dX^2 + dR^2 + R^2 d\Phi^2 + ds^2(\bE^6) \,,
\ee
with $\Phi\sim \Phi + 1$.
If we take the worldvolume coordinates to be
$(t,x,\varphi)$ with $\varphi \sim \varphi + 1$, then we may fix the
worldvolume diffeomorphisms for a D2-brane of cylindrical topology by
the `physical' gauge choice
\be
T=t\, ,\qquad X=x\, ,\qquad \Phi=\varphi\, .
\ee
For a cylindrical D2-brane of (possibly varying, but time-independent) radius 
$R(x,\varphi)$ at a fixed position in $\bE^6$ 
(and with the $X$-axis as the axis of symmetry) the induced metric is
\be
ds^2(g) = -dt^2 + dx^2 + R^2 d\varphi^2 + 
\left( R_x dx + R_\varphi d \varphi \right)^2 \,,
\ee 
where $R_x \equiv \pa_x R$ and $R_\varphi \equiv \pa_\varphi R$.
We will allow for a time-independent electric field $E$ in the
$x$-direction, and a time-independent magnetic field $B$, 
so the BI 2-form field strength is
\be
F= E \, dt\wedge dx + B \, dx\wedge d\varphi \, .
\label{fs}
\ee
Under these conditions the Lagrangian becomes
\be\label{lag}
{\cal L} = -\sqrt{(R^2+R_\varphi^2)(1-E^2) + B^2 + R^2 R_x^2} \,.
\ee
The corresponding Hamiltonian density is defined as
\be
{\cal H} \equiv \Pi E - {\cal L} \,,
\ee
where $\Pi \equiv \partial{\cal L}/\partial E$ is the `electric
displacement' subject to the Gauss law constraint $\partial_x\Pi=0$.

Let us now focus on the D2-brane supertube, for which 
$R$ is constant. In this case, the relation between the electric field
and the electric displacement takes the form
\be\label{elec}
E= \frac{\Pi}{R} \, \sqrt{B^2 + R^2 \over \Pi^2 + R^2}\, , 
\ee
so the Hamiltonian density becomes 
\be
{\cal H} = R^{-1} \, 
\sqrt{ \left( \Pi^2 + R^2 \right) \left( B^2 + R^2 \right)} \,, 
\ee
where it should be noted that $B= \bfna \times \bfa$ for BI 2-vector
potential $\bfa$. The Gauss law constraint implies that $\Pi$ is 
$x$-independent. In addition, the equation of motion for $\bfa$  
forces $B$ to be $x$-independent when $R$ is constant, as we are now
assuming.  Under these conditions ${\cal H}$ is $x$-independent, and its
integral over the circle parametrized by $\varphi$ yields a constant
energy per unit length, namely the tube tension 
\be
\tau = \oint d\varphi \, {\cal H} \,.
\ee
This is a function of $R$ and a functional of $\Pi(\varphi)$ and
$B(\varphi)$. For an appropriate choice of units, the integrals 
\be
q_s \equiv \oint d\varphi \, \Pi \qquad \mbox{and} \qquad 
q_0 \equiv \oint d\varphi \, B 
\ee
are, respectively, the conserved IIA string 
charge and the D0-brane charge per unit length carried by the tube. 
The total D0-brane charge is also conserved, so imposing periodic
boundary conditions on the tube, with period $L$, implies conservation
of $q_0$, and also that $q_0$ is quantized in multiples of some unit
proportional to $L^{-1}$. For fixed values of these charges, the
tube tension is bounded from below:
\be
\tau \geq |q_s| + |q_0| \,,
\ee
with equality if and only if 
\be
\label{bps}
\Pi = q_s \sac B = q_0 \sac R = \sqrt{|q_s q_0|} \,.
\ee
The crossed electric and magnetic fields generate a Poynting 
2-vector-density with 
\be
{\cal P}_\varphi = \Pi B 
\ee
as its only non-zero component. The integral of ${\cal P}_\varphi$ 
over $\varphi$ yields an angular momentum per unit length $J=\Pi B =q_sq_0$ 
along the axis of the cylinder.
It is this angular momentum that supports the tube at the
constant radius $\sqrt{|q_s q_0|}$. Substituting \eqn{bps} 
into (\ref{elec}) we see that 
\be
E = \mbox{sgn}(\Pi) = \pm1 \,.
\ee
This would be the `critical' electric field if the magnetic field  
were absent, as is shown by the fact that $\Delta = B$
when $|E|=1$.

\section{Supersymmetry}
We now aim to show that the tubular D2-brane configuration just described
is 1/4-supersymmetric; but as we also aim to relate it to some previously
discussed D2-brane configurations we shall now drop the assumption that
the worldvolume fields are independent of $x$ and $\varphi$.   
The number of supersymmetries preserved by any
D2-brane configuration is  the number of independent 
Killing spinors $\epsilon$ of the background for which
\be
\Gamma \epsilon =\epsilon \,,
\ee
where $\Gamma$ is the matrix appearing in the `$\kappa$-symmetry'
transformation of the worldvolume spinors \cite{kappa}. 
Introducing $\Gamma_\natural$
as the constant matrix with unit square which anticommutes
with all ten spacetime Dirac matrices, and 
$(\gamma_t,\gamma_x,\gamma_\varphi)$ as the
induced worldvolume Dirac matrices, we have \cite{BT}
\be
\Gamma = \Delta^{-1} \, \left(\gamma_{tx\varphi} + 
E \, \gamma_{\varphi} \Gamma_\natural + 
B \, \gamma_{t}\Gamma_\natural \right) \,.
\ee
For the D2-brane configuration of interest here 
\bea
&& \gamma_t = \Gamma_{\un{T}}\, ,\qquad  \gamma_x = \Gamma_{\un{X}} 
+ R_x \, \Gamma_{\un{R}} \,, \nn
&& \gamma_\varphi = R \, \Gamma_{\un{\Phi}}  + 
R_\varphi \, \Gamma_{\un{R}} \,,
\eea
where $\Gamma_{\un{T}},\Gamma_{\un{X}},\Gamma_{\un{R}}$ 
and $\Gamma_{\un{\Phi}}$ are the constant Minkowski spacetime 
Dirac matrices (with $\Gamma_{\un{\Phi}}^2=1$). 
For the spacetime coordinates that we have
chosen, any Killing spinor $\epsilon$ can be written as
\be
\epsilon = M_+ \, \epsilon_0 \sac 
M_\pm \equiv \exp\left( \pm {1\over2} \Phi \, \Gamma_{\un{R\Phi}} \right)
\,,
\ee
where $\epsilon_0$ is a constant 32-component spinor of $Spin(1,9)$.
The condition for preservation of supersymmetry can now be written as
\bea
0 &=& M_+  \, \left( R R_x \, \Gamma_{\un{TR\Phi}} 
+ B \, \Gamma_{\un{T}} \Gamma_\natural + 
R_x R_\varphi \, \Gamma_{\un{T}} - \Delta \right) \epsilon_0 \nn
&& + \, M_- \, \gamma_\varphi \Gamma_\natural \, 
\left( \Gamma_{\un{TX}} \Gamma_\natural + E \right) \epsilon_0 \,.
\label{susy}
\eea
It is clear that in order to satisfy this equation for all values of 
$\varphi$ (equal to $\Phi$ for our gauge choice) both terms on the
right hand side must
vanish independently. From the vanishing of the second term we recover
the condition that $E=\pm 1$, and we further deduce that $\epsilon_0$ 
must satisfy
\be\label{con1}
\Gamma_{\un{TX}} \Gamma_\natural \, \epsilon_0 = 
- \mbox{sgn}(E) \, \epsilon_0 \,.
\ee
The vanishing of the first term leads to $R_\varphi = 0$ and
\be
\left( R R_x \, \Gamma_{\un{TR\Phi}} + 
B \, \Gamma_{\un{T}} \Gamma_\natural \right)  \epsilon_0 = 
\sqrt{R^2 R_x^2 + B^2} \, \epsilon_0 \,.
\ee
For the supertube configuration $R_x=0$ and $B$ is constant, so the
constraint above becomes simply
\be\label{con2}
\Gamma_{\un{T}} \Gamma_\natural \, \epsilon_0 = 
\mbox{sgn}(B) \, \epsilon_0  \,.
\ee
The two conditions (\ref{con1}) and (\ref{con2}) are compatible and imply
preservation of 1/4 supersymmetry; the minimal energy tubular D2-brane
configuration is a {\it supertube}. Note that the constraints (\ref{con1})
and (\ref{con2}) are those associated with, respectively, 
a IIA superstring (in the $X$-direction) and D0-brane charge; 
in particular, there is no trace of the D2-brane in these conditions!
The physical reason for this is that a cylindrical D2-brane carries no
net D2-brane charge.

When $R_x \neq 0$, the vanishing of the first term on the right hand
side of \eqn{susy} implies that
\be\label{mag}
B = B_0 \, R R_x
\ee
for some constant $B_0$, and the constraint on the spinor becomes
\be\label{d2-d0}
\left( \Gamma_{\un{TR\Phi}} + 
B_0 \,\Gamma_{\un{T}}\Gamma_\natural \right)\epsilon_0 = 
\sqrt{1+ B_0^2} \, \epsilon_0 \,.
\ee
The Gauss law now implies that
\be
R(x) = C \, e^{x/E_0}
\label{Rx}
\ee
for some constants $C$ and $E_0$. This is just the
`BIon in a magnetic background' solution of \cite{GPTT} representing 
a IIA string ending on a bound state of D2-branes and D0-branes. 
To see this more clearly, first note that the constraint
\eqn{d2-d0} indeed corresponds to that of a D2-D0 bound state
in the $(R,\Phi)$-plane. Secondly, we can invert \eqn{Rx} to find
\be
x = E_0 \, \ln (R / C) \,,
\label{x}
\ee
which shows that $x$ is a harmonic function on the D2-brane
2-dimensional worldspace, as expected for the BIon \cite{CM,Gibbons}. 
Finally, using \eqn{x} we can rewrite the BI field strength \eqn{fs} as
\be
F = \frac{E_0}{R} \, dt \wedge dR + B_0 \, R dR \wedge d\varphi \,.
\ee
This corresponds to a radial Coulomb-like electric field on the D2-brane 
worldspace, as expected from the charge at the endpoint of the string, 
and to a constant density of D0-brane charge per unit worldspace area,
precisely as in the solution of \cite{GPTT}.

\section{Discussion}
By definition, a Dp-brane is a sink for IIA string charge. However,
the D0-brane, being point-like, 
is special because the IIA charge has nowhere to go and must
exit on another IIA superstring. Thus D0-branes can appear as `beads'
on a IIA superstring, breaking the 1/2 supersymmetry of the string to
1/4 supersymmetry. Of course, quantum mechanics will ensure that
the ground state of such a superstring is one for which the D0-brane
charge is uniformly distributed along the string. This `charged' IIA
superstring will have a tension exactly equal to the D2-brane
supertube discussed above, $\tau = |q_s| + |q_0|$. What 
distinguishes them is the angular momentum; the charged superstring 
has zero angular momentum while the supertube has angular momentum
per unit length $J$ equal to $q_s q_0$. In principle, $J$
can be specified {\it independently} of the string and D0-brane
charges, so given $q_s$ and $q_0$ we might expect there to be some 
supersymmetric string/tube configuration with arbitrary $J$. 
As long as $|J|\le |q_s q_0|$ it is not difficult to see what this will
be: a supertube with angular momentum per unit length 
$J$, together with a charged superstring along its central axis. 
Because $J$ is quantized in the same units as $B$, it is always possible
for excess string and D0-charge to `condense' out of a tubular
D2-brane to leave behind a supertube supported from collapse 
by any given $|J|$ less than $|q_s q_0|$. On the other hand, it is
unclear what the ground state could be when $|J| > |q_s q_0|$. It is
conceivable that $|q_s q_0|$ is an upper bound on the angular momentum of
a
{\it supersymmetric} IIA superstring configuration with charges $q_s$
and $q_0$, and that supersymmetry is spontaneously broken when 
$|J|>|q_s q_0|$. 

There is some similarity here to the status of angular momentum in the
context of black holes of $D=5$ supergravity \cite{blackholes,GMT,GH}. 
The black hole mass is determined entirely by its charge, if it is
supersymmetric. For a given charge there is a one-parameter 
family of supersymmetric black hole spacetimes, parametrized by an 
angular momentum, but there is a critical value of the angular
momentum beyond which the physics is qualitatively 
different. There is also a similarity to the suggestion \cite{goliath}
that the `giant graviton' ground state is not supersymmetric above a
critical value of the angular momentum, although the 
supersymmetry of relevance in that case is $AdS$ rather than 
Poincar\'e.
 
Finally, we wish to comment on some M-theory configurations dual to the
D2-brane supertube. The first one involves the M2-brane: 
the D2-brane action is equivalent to the action for the $D=11$
supermembrane, the equivalence involving an exchange of 
the BI 1-form potential for a periodically-identified scalar field
$\Psi$ (with unit period) representing position in the 11th
dimension \cite{BT}. Specifically, one has
\be
\partial_i \Psi = {g_{ij}\varepsilon^{jkl}F_{kl}\over 
2 \sqrt{-\det (g +F)} } \,.
\ee
{}For the D2-brane supertube this yields 
\be
\Psi = \Pi \, \varphi - t \,.
\ee
This implies, firstly, that the M2-brane is wound $\Pi$ times around
the 11th dimension, as expected given the identification of $\Pi$ with
IIA string charge. Secondly, it implies that $\Pi$ must be an integer,
obviously equal to the number of IIA strings dissolved in the original
D2-brane. Thirdly, it implies that there is a wave at the speed of
light in the 11th dimension, as expected from the D0-brane charge; the
momentum of this wave is proportional to $B$. Note that if 
the $x$ dimension is periodically identified with period $L$ then 
one can take $L\rightarrow 0$ to arrive at a new (1/4)-supersymmetric
IIA configuration in which a helical string rotates, producing a net
momentum along the axis of the helix \cite{KM}. 

Another dual M-theory configuration consists of an
M5-brane with topology $\bR^4 \times S^1$, carrying dissolved membrane
charges oriented along orthogonal planes in $\bR^4$. To see this, 
let the $S^1$ be parametrized by $\varphi$, and let the two planes 
span the directions
1-2 and 3-4. Reducing to the IIA theory by compactifying the
4-direction, and T-dualizing along 2 and 3 leads to the D2-brane
supertube.

\medskip
\section*{Acknowledgments}
\noindent
We thank Barak Kol for correspondence. D.M. is supported by a PPARC fellowship.



\begin{thebibliography}{99}
\bibitem{footnote}
This is an
oversimplification because the integral of the charge density of the
cylindrically-symmetric solution has an IR logarithmic divergence; for
this reason the `charge' $Q$ mentioned below should be regarded only as a
measure of scale.
\bibitem{KM}
We owe this observation to Barak Kol and Juan Maldacena; note that
T-duality along the axis of the helix interchanges the string with the
wave yielding a IIB version of the rotating helical string.
\end{thebibliography}
\end{document}